\documentclass[num-refs]{wiley-article}

\usepackage{siunitx}
\usepackage{graphicx}
\usepackage[version=3]{mhchem}
\usepackage{amssymb}
\usepackage{amsmath}
\usepackage{booktabs}
\usepackage{multirow}
\usepackage{tabularx}
\usepackage{mciteplus}
\usepackage{caption, subcaption}
\usepackage{afterpage}
\usepackage{float}
\usepackage{lineno}

\papertype{Communication}

\title{Theoretical design of effective multilayer optical coatings using oxyhydride thin films}

\author[1]{E. Strugovshchikov}
\author[1]{A. Pishtshev}
\author[2]{S. Karazhanov}

\affil[1]{Institute of Physics, University of Tartu, Tartu, Estonia}
\affil[2]{Department for Solar Energy, Institute for Energy Technology, Kjeller, Norway}

\corraddress{Dr.~Aleksandr~Pishtshev, 
Institute~of~Physics, University~of~Tartu, 50411 Tartu, Estonia}
\corremail{aleksandr.pishtshev@ut.ee}

\begin{document}

\begin{frontmatter}
\maketitle

\begin{abstract}
Rare-earth metal oxyhydride compositions are currently attracting increasing attention to develop materials with unusual optical responses. In this study, using computer simulations of the electronic and optical properties we studied the optical responses of two stable yttrium oxyhydride compounds, \ce{Y4H10O} and \ce{YHO}, for the visible light range. The emphasis was on modeling macroscopic optical characteristics, which we numerically derived within a conventional scheme using refractive indices, and absorption, transmittance, and reflection spectra. Our main goal was twofold: first, to simulate spectral behavior of different single-phase and two-phase oxyhydride compositions, and second, to perform a comparative analysis that could explain the features of the transmission spectra measured for different samples. Based on the obtained results, we proposed models of new optical coatings in which yttrium oxyhydrides play the key role. 
In the context of nonlinear optics, we evaluated the frequency profile of the second-order susceptibility ${\chi}^{(2)}(2\omega)$ for the noncentrosymmetric cubic structure of \ce{Y4H10O} and showed that this system could exhibit large optical nonlinearity. 
\end{abstract}
\end{frontmatter}

\section{\label{sec:level1} Introduction }
An oxyhydride is a complex inorganic chemical compound in which the rare earth or transition metal is a cation that is balanced by two adjacent anions, \ce{O^{2-}} and \ce{H^{-}}, forming different compositional ratios~\cite{mixed2,Kageyama2018,Pishtshev2019}.   
The development of oxyhydrides began more than 20 years ago~\cite{Fokin1995,Fokin1996,Hayward1882,Fokin2004} as an effort to improve some key properties of transition metal hydrides.
It was found that the direct addition of oxygen to the hydride lattice not only varies the stoichiometry by stabilizing the oxygen positions in the crystal lattice, but also expands the possibilities of chemical control of the structural and electronic properties of such a modified metal-hydride system~\cite{Fokin1995,Fokin1996,Hayward1882,Fokin2004,oxyhydrides,oxyhydrides2,Pishtshev2014}.
The effect of partial oxidation makes the variation of the crystal structure stable, since chemical interactions form a reconfiguration of the valence charges that provides a new minimum of free energy~\cite{Pishtshev2019,Pishtshev2014}.
Compared to the initial metal hydride, the ground state of the oxyhydride system exhibits better compositional and structural stability; its macroscopic mechanical properties are similar to those of complex crystalline oxides.
Several yttrium oxyhydrides of certain compositions exhibit unusual properties that are very attractive for use in developing optical coatings. For this reason, yttrium oxyhydrides as photochromic compounds have gained increased interest.
A large number of experimental works have focused on the study of optical and electronic properties. In particular, we can mention the reports on the discovery of persistent photoconductivity and reversible color change under the influence of light on \ce{YHO} thin films~\cite{MONGSTAD2011}.
In this work, the photochromic darkening effect was confirmed by observing optical transmittance when the oxyhydride film was illuminated by light incident at different angles.
The photochromic response, electronic properties, and vacancy formation in yttrium oxyhydrides have been examined in works~\cite{Nafezarefi2017,Plokker2018}.
Interesting results were obtained when analyzing the photochromic compositions of oxyhydride thin films~\cite{Moldarev2018,Moldarev2018_2}.
Characteristics of the electronic structure were evaluated from UV-vis spectrophotometry~\cite{MOLDAREV2020} and optical ellipsometry~\cite{MONTERO2018} studies.

Given that the integration of the anionic degrees of freedom \ce{O^{2-}} leads to a new functionality, it is important to understand how a chemical composition of a condensed oxyhydride compound in association with a particular crystal structure affects its ability to respond to light illumination.
Analysis of the structural and charge characteristics showed that in yttrium and some rare-earth oxyhydrides, anions form a subsystem of highly localized charges~\cite{Pishtshev2019}, what may be the key factor affecting electronic and optical responses.
This means that the photochromic properties of thin films depend on how valence electrons can effectively interact with structural and charge configurations belonging to different yttrium oxyhydride compositions~\cite{Pishtshev2014}.
However, the lack of detail information on the mechanisms of the structure-property relationships causes a number of difficulties in the conceptual interpretation of experimental data.
In particular, many aspects of the photochromic effect in oxyhydrides are not fully understood. For example, the mechanism of how ultraviolet/visual light illumination stimulates the corresponding optical responses, such as the reversible color change of the oxyhydride film, remains still unknown~\cite{Montero2017,Baba2020_2}.
Thus, the goal of the present study is twofold:
First, to overcome the lack of crystal-chemical information in the experimental work and, therefore, to model the problem in such a way that it can be effectively addressed, we will consider two chemical models-the low-oxidized and the high-oxidized forms of yttrium hydride. That is, based on the results of the density functional theory (DFT) calculations, we will perform the standard theoretical work on the modeling of the optical properties of both systems by evaluating the spectral behavior of the bulk optical characteristics. This will allow a comparison with the experiment to verify the correctness of the model functions and numerical parameters when reproducing the experimental data.
Accordingly, for two different cases, low- and highly-oxidized yttrium hydrides, the spectral dependencies of refractive index, reflection and transmittance will be discussed and presented in terms of the macroscopic electronic dielectric function. 
Second, the data obtained will be used to efficiently model the optical behavior that structural assemblies consisting of combinations of different films can exhibit.
This modeling approach allows us to select potential candidates that can serve as reference layers for optical coatings made from multilayer structures.
Its important advantage is the possibility to effectively combine different oxyhydride phases in order to model thin films that could have such desirable properties as high transparency, anti-reflectivity, and sensitivity,~{\it etc}.

\section{\label{sec:level2}Simulation details, methods and modeling}

The Vienna ab initio Simulation Package (VASP)~\cite{vasp1,vasp2} and
projector augmented-wave (PAW) method \cite{bloh,paw} were used for
simulations of the electronic structure and optical properties. 
The periodic DFT calculations employed PAW pseudo-potentials and the plane-wave 
basis sets taken as $4s^2 4p^6 5s^2 4d^1$, $2s^2 2p^4$ and $1s^1$ 
for \ce{Y}, \ce{O} and \ce{H} elements, respectively.
Non-local exchange effects have been taken into account by using the range-separated hybrid Heyd-Scuseria-Ernzerhof (HSE06) functional~\cite{hse1,hse2,hse3} 
The fraction of the Fock exchange was estimated in the scheme of an ion–covalent 
semiconducting system~\cite{Audrius2008,Vidal2010,Audrius2011,Mad2011,He2012} 
via the evaluation of the inverse dielectric constant  
$\epsilon_{\infty}^{-1}\,$. These calculations, which have taken into account local field effects, were performed in accordance with the procedure of the density functional perturbation theory as implemented in VASP \cite{Gajdo2006}.
The $\Gamma$-centered optimized $8\times8\times8$ ${\bf k}$-points grid
and $700$ eV cutoff energy have been used for the numerical integration over 
a Brillouin zone (BZ).
Macroscopic optical properties were evaluated on the base 
of the frequency-dependent complex dielectric function 
$\epsilon(\omega) = \epsilon_1(\omega) + i\epsilon_2(\omega)$.
The IMD software~\cite{IMD} was used to simulate optical coatings consisting of multilayer combinations.
The Elk code~\cite{elk} was used to calculate the frequency profile of the 2nd-order susceptibility.

\begin{table}[!htbp]
\centering
\caption{\label{tab:crystal} Summary of crystallographic data for cubic and orthorhombic structural models of two chemical compositions of bulk yttrium oxyhydride. The following characteristics are shown: lattice parameters $a/b/c$, unit cell volume $V$, density $\rho$, theoretical hardness $H_V$, shortest cation-anion distances \ce{Y-H} and \ce{Y-O}, and reaction enthalpy $\Delta H$ estimated for the decomposition into two binary compounds \ce{YH2} and \ce{Y2O3}.}
\begin{tabular}{lclccccccc}
\headrow
Chem. & Ref. & Cryst. & $a/b/c$ & $V$ & $\rho$ & $H_V$ & \ce{Y-H} & \ce{Y-O} & $\Delta H$\\
\hline
formula & & struct. & (\AA) &  (\AA$^3$) & (g/cm$^3$)& (GPa) & (\AA) & (\AA) & (kJ/mol) \\
\hline
 \hiderowcolors
\ce{Y4H10O} & Theory~\cite{Pishtshev2019} & $P{\overline{4}}3m$ & $5.23$ &  $143.03$ & $4.43$ & $8.54$ & $2.26$ & $2.29$ & $-250.2$ \\
            & Exp.~\cite{MAEHLEN2013} & $P{\overline{4}}3m$ & $5.24$ &  $143.88$ &  &  & $-$ & $-$ &  \\
 \cmidrule{1-9}
\ce{YHO}& Theory~\cite{Pishtshev2019} & $F{\overline{4}}3m$ & $5.29$ & $148.20$ & $4.75$ & $8.71$ & $2.29$ & $2.29$ & $-33.1$ \\
            & Exp.~\cite{MONTERO2018} & $F{\overline{4}}3m$ & $5.29$ &  $148.03$ &  &  & $-$ & $-$ &  \\
 \cmidrule{2-9}
 & Theory~\cite{Pishtshev2019} & $Pnma$ & $7.54$, $3.77$, $5.33$ &  $151.29$ & $4.65$ & $7.99$ & $2.31$ & $2.24$ & $-31.9$ \\
            & Exp.~\cite{yho62} & $Pnma$ & $7.54$, $3.76$, $5.33$ &  $150.83$ &  &  & $2.31$ & $2.22$ &  \\
 \hline
\end{tabular}
\end{table}
Comparison of the experimental studies~\cite{Montero2017,Baba2020,yho62} with the corresponding theoretical results~\cite{Pishtshev2014,Pishtshev2019} showed that the crystal architecture of the synthesized yttrium oxyhydrides can be modeled as follows: (i) a low-oxidized \ce{Y4H10O} composition with a cubic phase in the $P{\overline{4}}3m$ space group, and (ii) a high-oxidized \ce{YHO} composition with two structures, a cubic phase in the $F{\overline{4}}3m$ space group and an orthorhombic phase in the $Pnma$ space group. All these compositions are covered by the stability region of the phase diagram ~\cite{Pishtshev2019}. 
Table~\ref{tab:crystal} lists the crystallographic data of these models. These data were used to simulate the electronic structure and optical properties of \ce{Y4H10O} and \ce{YHO}. 
\begin{table}[!htbp]
\centering
\caption{\label{tab:eps} Some optical and electronic characteristics evaluated for the models of Table~\ref{tab:crystal}; $\epsilon_{\infty}$ - electronic dielectric constant,
$n$ - the refractive index, $E_g^{opt}$~-~the value of optical band gap.
(For comparison with ion-covalent wide-gap oxide \ce{Y2O3}~\cite{Nigara_1968,Nigara1971}: 
$E_g^{opt}\,=\,5.6$ eV, $\epsilon_{\infty}\,=\,3.58$ and $n\,=\,1.93$.)}
\begin{tabular}{llcccl}
\headrow
Chem. & Cryst. & $\epsilon_{\infty}$ & $n$ & $E_g^{opt}$ & Character of\\
\hline
formula & struct. & & & eV & the band gap \\
\hline
 \hiderowcolors
\ce{Y4H10O} & $P{\overline{4}}3m$ & $7.87$ & $2.81$ & $2.2$ & direct  ($R$)\\
 \cmidrule{1-6}
\ce{YHO}& $F{\overline{4}}3m$ & $4.75$ & $2.18$ & $4.9$ & indirect ($
X$-$\Gamma/M$)\\
 \cmidrule{2-6}
 & $Pnma$ & $4.71$ & $2.17$ & $4.3$ & indirect ($Z$-$Y$) \\
 \hline
\end{tabular}
\end{table}
For the given models of yttrium oxyhydrides Table~\ref{tab:eps} specifies the electronic dielectric constants $\epsilon_{\infty}$ (together with the estimate of the refractive index) as well as describes two key parameters of light absorption in the optical spectral range: absorption edge (optical band gap $E_g^{opt}$) and character of the gap (associated with the relevant point in BZ).

\section{\label{sec:level3}Results and discussion}
The effect of partial oxidation, which causes stoichiometric changes in the anion sublattice of yttrium hydride, directly affects the structural characteristics of the crystal lattice. In the present work we illustrate how this modifies the structure-optical properties relationships. The theoretical study of the electronic structure by means of the hybrid HSE06 functional for several structural models of yttrium oxyhydrides is reflected in Supplementary Figures~S1$-$S8, where we have presented site-projected and total density of states (DOS), band structure along with the spectral behavior of the absorption coefficient. The characterization of the optical properties is given in Figure~\ref{fig:Comb_optics}, which shows the spectral dependence of the dielectric function ($\epsilon$), refractive index ($n$), reflection ($R$) and transmittance ($T$) spectra.

The electronic band structure of the bulk \ce{Y4H10O} is characterized by
the direct band gap located in the $R$ point ($0.5;0.5;0.5$) of BZ. 
The topmost part of the valence band (VB) has a diffuse structure represented mainly by $1s$-orbitals of hydrogen. The lowest empty $3d$-orbitals states of yttrium form the lower part of the conduction band (CB). The $2p$ oxygen states, which are strongly hybridized with the $1s$ oxygen states, are red-shifted in energy by $1$ and $3$ eV from the top of the VB.  In light of possible electronic transitions, the spectral behavior of the $\alpha$ absorption coefficient, shown in Supplementary Figure~S9, agrees well with the site-projected DOS picture. It can also be seen that in \ce{Y4H10O} the absorption edge lies for photon energies around $2.2$ eV.
After the addition of a large amount of oxygen, the strong covalence of the $\ce{Y-H}$ connections is modified by the contribution of a significant fraction of the ionicity from the new $\ce{Y-O}$ connections. As it follows from Supplementary Figures~S1$-$S4,  
an increase of the oxygen content leads to the significant enlargement of 
the dielectric gap for all possible phases of \ce{YHO}.
Moreover, there exists the specific trend showing the interplay between lattice and electronic degrees of order in the highly-oxidized composition \ce{YHO}.
The most stable cubic phase $F{\overline{4}}3m$ possesses the indirect band gap 
of $4.9$ eV which is largest in comparison with the $P{\overline{4}}3m$ cubic, 
$4.1$ eV, and the $Pnma$ orthorhombic, $4.3$ eV, geometries.
The $P{\overline{4}}3m$ cubic modification exhibits 
the direct band gap located in the $X$ point ($0;0.5;0$) of BZ.
In the orthorhombic structure, the dielectric gap becomes again indirect.
On the other hand, as the oxygen content increases, the hybridization between the $2p$ oxygen and $1s$ hydrogen states becomes stronger. The $sp$-hybridization effect stretches the mixing of these states down more than $-4$ eV from the top of the VB. The lowest part of the CB is characterized by mixing between the $3p$ oxygen and $4d$ yttrium empty states, in which the contribution of the oxygen states is negligible.
In the context of confirming our theoretical calculations, the analysis shows that the calculated DOS given in Supplementary Figures~S1$-$S4 are in good agreement with the measured XPS spectra for thin films of chemical composition close to the formula \ce{YHO}~\cite{Baba2020}.
Supplementary Figures~S10$-$S12 presents the comparison
how the different structural phases of \ce{YHO} determine 
the various features of optical absorption.
Here it is interesting to note a common feature of the absorption spectra ($\alpha({\omega})$): their increase is noticeably limited.
This is a direct consequence of the strong Coulomb repulsion
between valence shells of different anions. Near the top of VB 
the repulsion creates a flat region of low DOS intensity with 
such a devastating effect on performance of the electron transitions.
Therefore, the increase in absorption becomes significant only beginning from photon energies of $5.5$ eV and higher, i.e., when electronic transitions between $2p$ states of oxygen and $4d$ states of yttrium become dominant.

Figure~\ref{fig:Comb_optics} analysis helps to understand what opportunities for optical coating design can be realized with yttrium oxyhydrides.
First, we can see how the features of interband electronic transitions are reflected in the optical properties of oxyhydrides with low and high oxygen content.
The main difference between them is the blue shift of the photon absorption edge into the UV region, observed for all phases of the \ce{YHO} system. We have already highlighted this effect in Supplementary Figures~S9$-$S12; it is caused by a change of the band gap when the oxygen content increases (Table~\ref{tab:eps}). Accordingly, other optical characteristics are also affected by the blue shift factor. On the other hand, as can be seen from Figure~\ref{fig:Comb_optics}, in the spectral region relating to
visible light the cubic phase of \ce{Y4H10O} exhibits a larger refractive index 
than that of any phase of \ce{YHO}.

\begin{figure}
\begin{subfigure}{.32\textwidth}
\centering
     \includegraphics[width=0.80\textwidth]{Figure_02a.pdf}
\caption{\label{fig:Fig_2a} Schematic view of the bi-phase composite system 
consisting of \ce{YHO} ($89\%$) and \ce{Y4H10O} ($11\%$). Layer thicknesses are given in nanometers.}
\end{subfigure}\hfill
\begin{subfigure}{.32\textwidth}
\centering
     \includegraphics[width=0.98\textwidth]{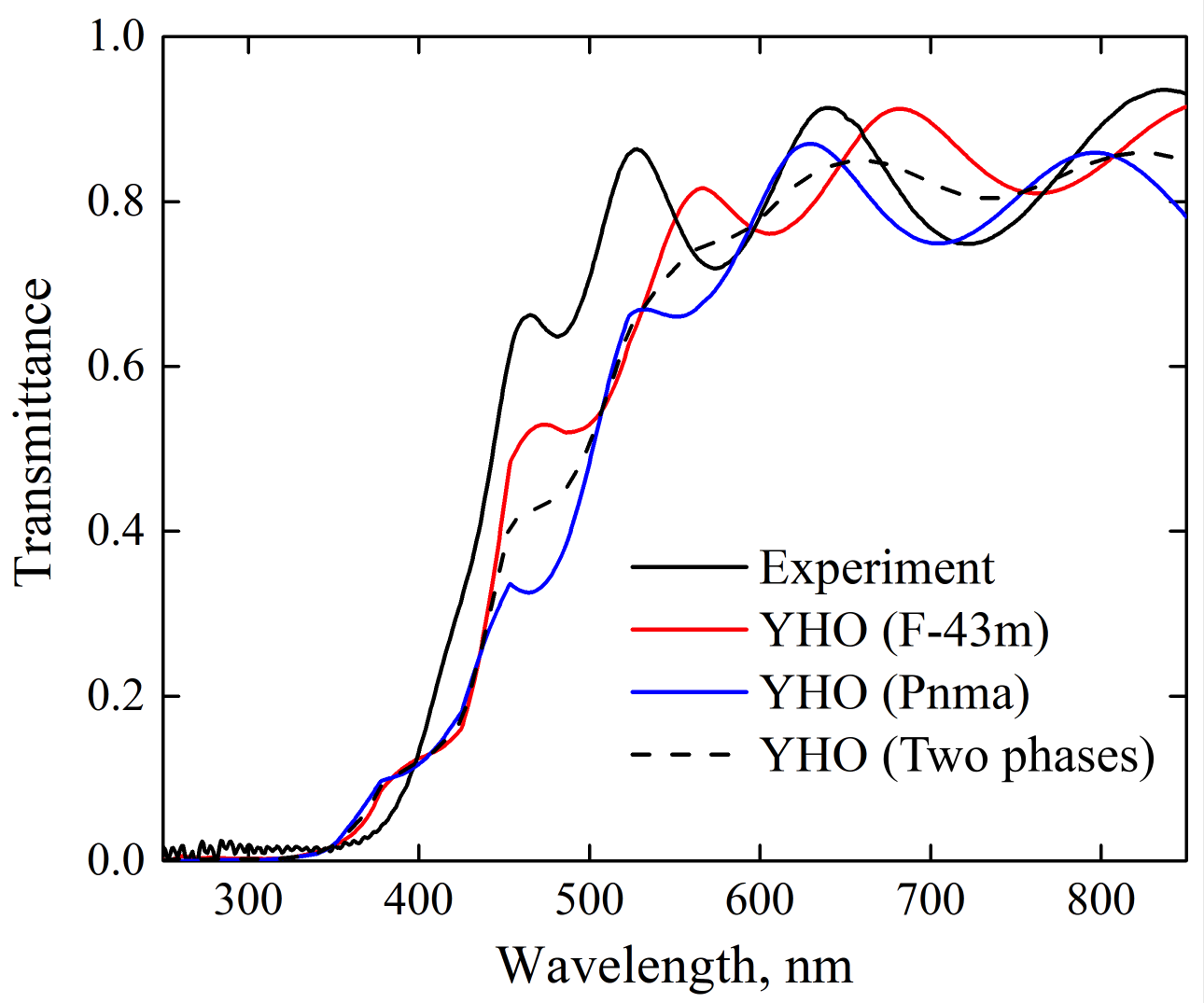}
\caption{\label{fig:Fig_2b} Experiment~\cite{Baba2020} -- black line, theory -- red, blue and dashed lines.}
\end{subfigure}\hfill
\begin{subfigure}{.32\textwidth}
\centering
     \includegraphics[width=0.98\textwidth]{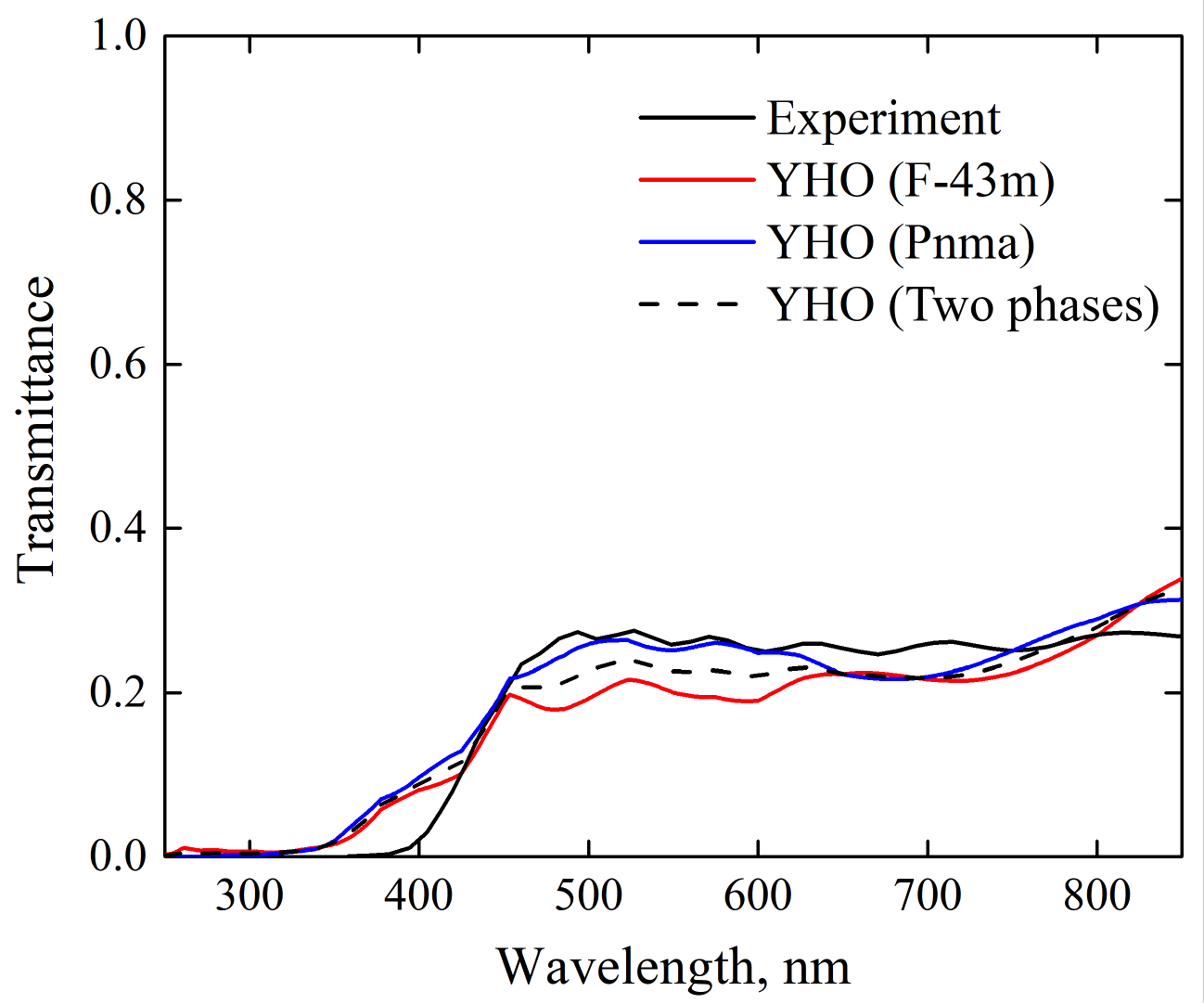}
\caption{\label{fig:Fig_2c} Experiment~\cite{Baba2020} -- black line, theory -- red, blue and dashed lines.}
\end{subfigure}\hfill
\caption{\label{fig:model} Design and optical properties of the proposed bi-phase composite system. Experimental work and measurement data are presented in Ref.~\cite{Baba2020}. Part (b) -- transmittance spectra for the transparent state, Part (c) -- transmittance spectra for the photodarkened (opaque) state.}
\end{figure}

Second, the use of accurate DFT calculations allows one to analyze the experimental transmission spectra through a detailed comparison with simulation results. The goal is, by finding spectral dependences similar to the experimental ones, to select a set of suitable chemical configurations, which can then be used to build the most relevant structural model.
To prove the effectiveness of such theoretical work, we simulated the optical response of the two-phase composite system, which, as shown in Figure~\ref{fig:Fig_2a}, represents a layered model consisting of \ce{Y4H10O} and \ce{YHO} thin films.
The best fit of the simulation data to the experiment was chosen from the numerical variation of the (\ce{Y4H10O}){$\,:\,$}({\ce{YHO}}) relative content.
Shown in Figure~\ref{fig:Fig_2a} is a proposed model with spatial distribution of layers comprised of $11\%$ of \ce{Y4H10O} and $89\%$ of \ce{YHO}. 
The transmission spectra for transparent and photodarkened states calculated for such system are shown in Figures~\ref{fig:Fig_2b} and~\ref{fig:Fig_2c}, respectively. 
For structural modeling of \ce{YHO}, three options were taken into account: separately cubic and orthorhombic phases, as well as their combination in equal proportions.
For comparison, we also presented experimentally-measured transmission spectra for a synthesized sample of heterogeneous yttrium oxyhydride film~\cite{Baba2020_2}. 
Clearly, the obtained results help us not only to understand morphological and structural features of the formation and functionalization of oxyhydride films (including the effect of darkening during light irradiation in a heterogeneous structure), but also to overcome the difficulties~\cite{Baba2020} associated with chemical processes of partial oxidation.

\begin{figure}
\begin{subfigure}{.32\textwidth}
\centering
     \includegraphics[width=0.80\textwidth]{Figure_03a.pdf}
\caption{\label{fig:Fig_3a} Schematic view of the proposed model in a cross-section of layers. Layer thicknesses are given in nanometers.}
\end{subfigure}\hfill
\begin{subfigure}{.32\textwidth}
\centering
     \includegraphics[width=0.98\textwidth]{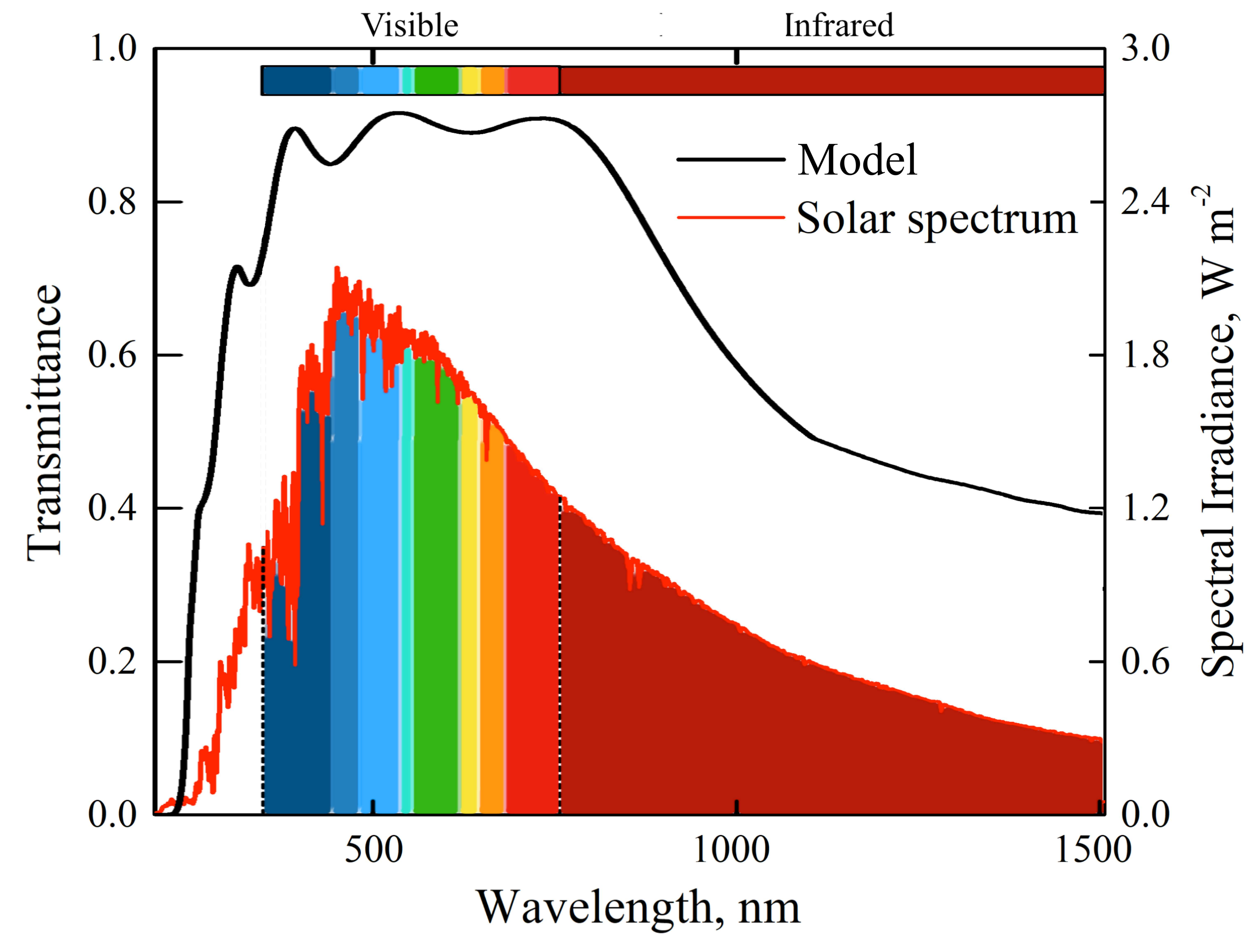}
\caption{\label{fig:Fig_3b} Predicted transmittance spectrum {\it vs} available solar radiation. The solar spectrum was drawn on the base of solar irradiance data of Ref.~\cite{Hulstrom1985}. }
\end{subfigure}\hfill
\begin{subfigure}{.32\textwidth}
\centering
     \includegraphics[width=0.98\textwidth]{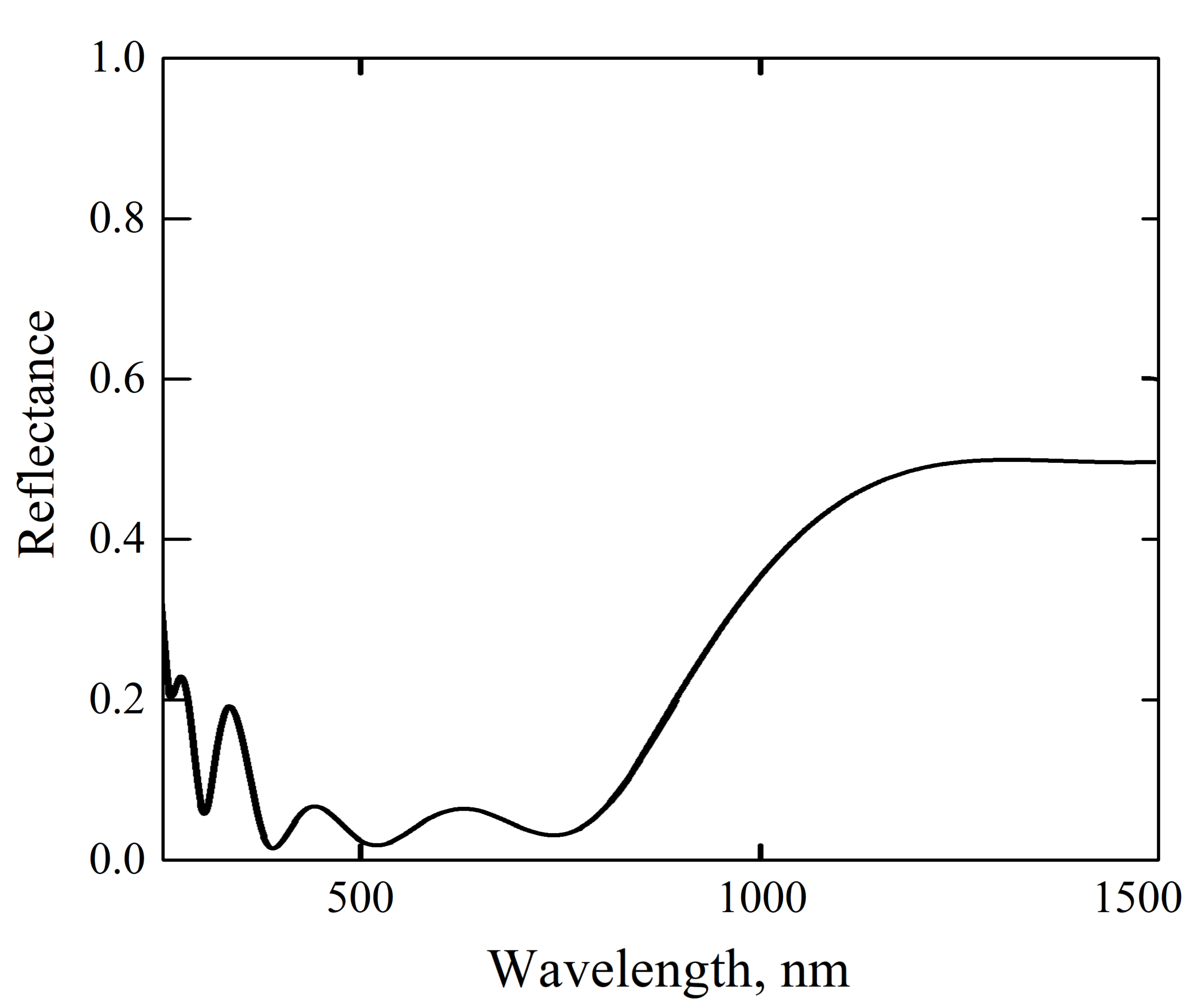}
\caption{\label{fig:Fig_3c} Predicted reflectance spectrum. }
\end{subfigure}\hfill
\caption{\label{fig:solar_trans} 
Design and optical properties of an oxyhydride model of a low-emissivity coating.}
\end{figure}

\begin{figure}
\begin{subfigure}{.32\textwidth}
\centering
     \includegraphics[width=0.70\textwidth]{Figure_04a.pdf}
\caption{\label{fig:Fig_4a} Schematic view of the proposed model in a cross-section of layers. Layer thicknesses are given in nanometers.}
\end{subfigure}\hfill
\begin{subfigure}{.32\textwidth}
\centering
     \includegraphics[width=0.98\textwidth]{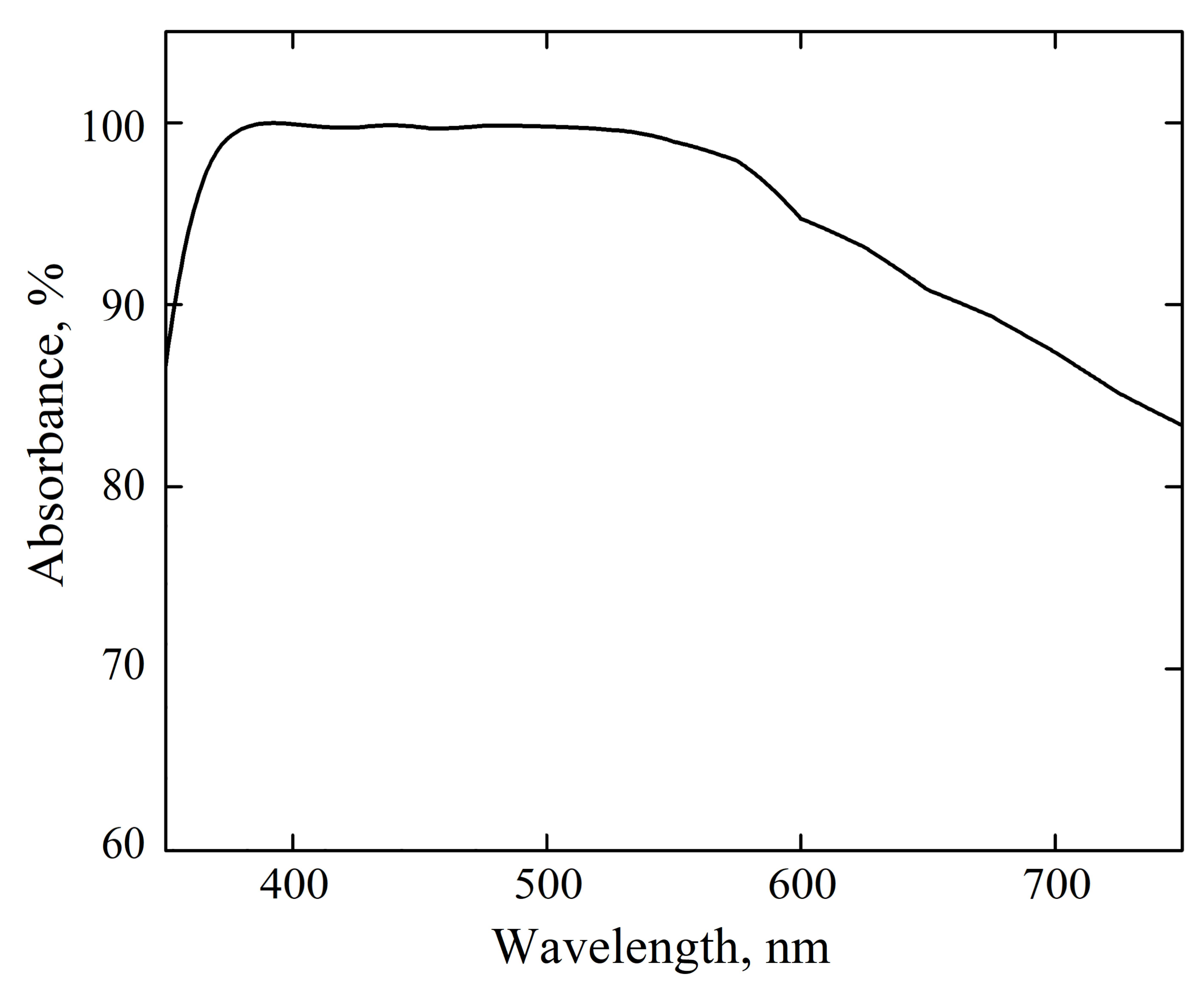}
\caption{\label{fig:Fig_4b} Predicted absorbance spectrum. }
\end{subfigure}\hfill
\begin{subfigure}{.32\textwidth}
\centering
     \includegraphics[width=0.98\textwidth]{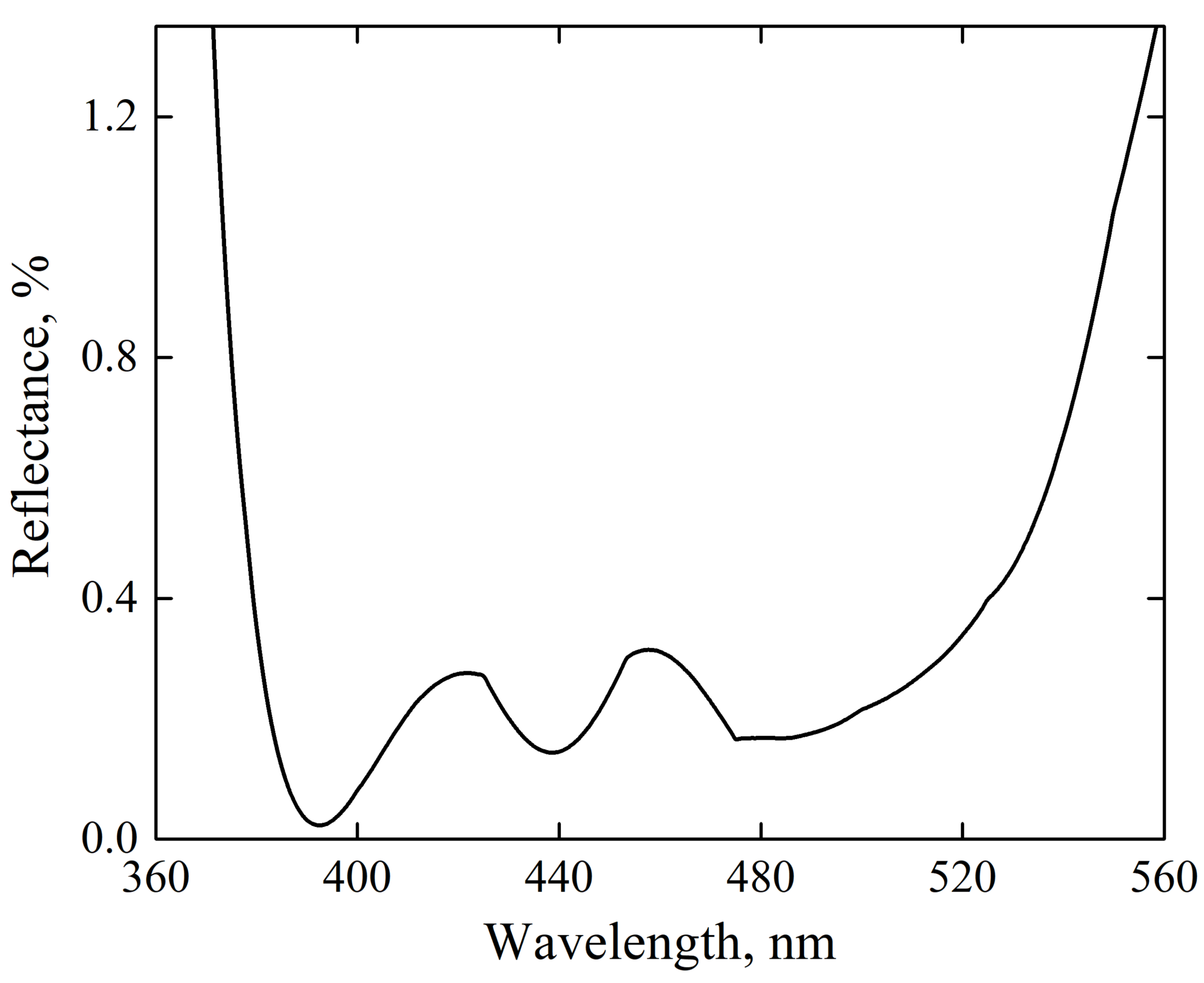}
\caption{\label{fig:Fig_4c} Predicted reflectance spectrum. }
\end{subfigure}\hfill
\caption{\label{fig:absorb_model}
Design and optical properties of an oxyhydride model of a light-absorbing non-reflective coating.}
\end{figure}

\begin{figure*}[b]
\centering
     \includegraphics[width=0.82\textwidth]{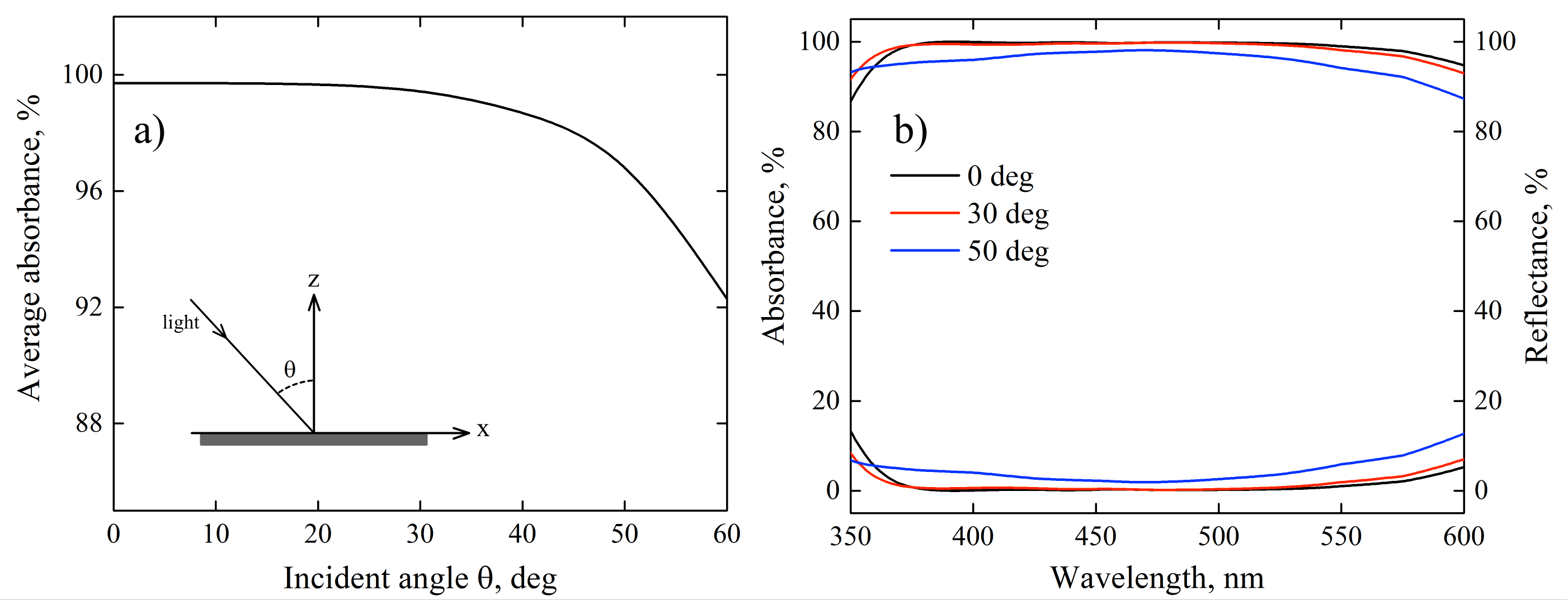}
\caption{\label{fig:Fig_5}
a) Angle dependence of the optical absorption for the oxyhydride model of the light-absorbing non-reflective coating. The absorption values were averaged over the $375$-$550$ nm wavelength range.
b) Absorbance and reflectance spectra calculated for three angles of incident light: $0$, $30$ and $50$ degrees. }
\end{figure*}

Also of interest from a practical point of view is the compositional and structural variability of oxyhydrides. Since the frequency-dependent reflection spectra $R$ and transmittance spectra $T$ for different yttrium oxyhydride phases differ from each other, this opens good opportunities to create a prototype multilayer composite optical system that could combine various phases and oxyhydride compositions.
To demonstrate how the combination of the effects of refraction, transmittance, and reflection together with the effects of internal interference of light waves can be utilized, we proposed two different prototypes of optical coatings in which oxyhydride thin films play a key role.\\
(i) {\it Model of a low-emissivity coating.} 
The model is expected to block ultraviolet (UV) radiation and, to a specified extent, heating caused by infrared (IR) solar radiation; the model efficiency is that visible light can propagate with high transmittance (above $90$\%) and low reflectance (below $10$\%).
Design details are summarized in Figure~\ref{fig:Fig_3a}. 
Wide-gap \ce{Y2O3}, which is used for optical coatings in a broad spectral range from the near-UV to the IR region, was chosen as the material for the upper and lower layers.
The intermediate layers are represented by two films: a nanoscale silver layer, suitable for increasing near-IR reflectivity, and a thin layer based on \ce{YHO}.
The key role of the oxyhydride layer is to regulate the transmission and reflection of light depending on the frequencies of the spectral range.
The substrate was approximated by amorphous silicon glass (\ce{a-SiO2}).
Due to the proposed "low to high to low" light transmission scheme (Figures~\ref{fig:Fig_3b} and~\ref{fig:Fig_3c}), the suggested design reduces both UV and IR light radiations.
As shown in Figure~\ref{fig:Fig_3b}, the transmittance is excellent over the entire visible spectral range. The performance of the model is very similar to that of existing analogues: the solar heat gain coefficient is about $40$\%, light-to-solar gain ratio is $2.25$, and the transmittance window ($T>80$\%), being between $360$ and $855$~nm, provides an average visible light transmittance of $90$\%. \\
(ii) {\it Model of a light-absorbing non-reflective coating.} 
The model is expected to exhibit high light absorption and low light reflection.
Design details are summarized in Figure~\ref{fig:Fig_4a}. As in the former case, \ce{Y2O3} serves as the top cover layer, \ce{a-SiO2} is the substrate layer. 
The metal-dielectric interfaces, which are represented by a combination of yttrium and oxyhydride thin films, are the main elements of the model; their central function is to significantly improve the light-absorbing properties~\cite{Craighead1979}. 
To successfully predict a composite model that would be close to a material with "ideal" absorption levels, the main goal of the simulation was to find the proper match between the thicknesses of several layers. 
Figure~\ref{fig:Fig_4a} illustrates the optimal selection of different layers. 
Figures~\ref{fig:Fig_4b} and~\ref{fig:Fig_4c} present calculations for the optical properties of this system; Figure~\ref{fig:Fig_5} describes how incident light waves are efficiently absorbing from a large range of angles.
The results demonstrate that the proposed model can serve as the basis for a broadband absorption coating. Over a specific wavelength range of $375-550$~nm, in the case of normal incidence, the system absorbs on average $99.7$\% of incoming light radiation; the highest level of absorption, $99.98$\%, is reached at $395$~nm. However, at longer wavelengths of visible light, the system loses its feature as almost perfect absorbing as its absorption gradually decreases to $90$\%. %

\begin{figure}[!htbp]
\centering
     \includegraphics[width=0.48\textwidth]{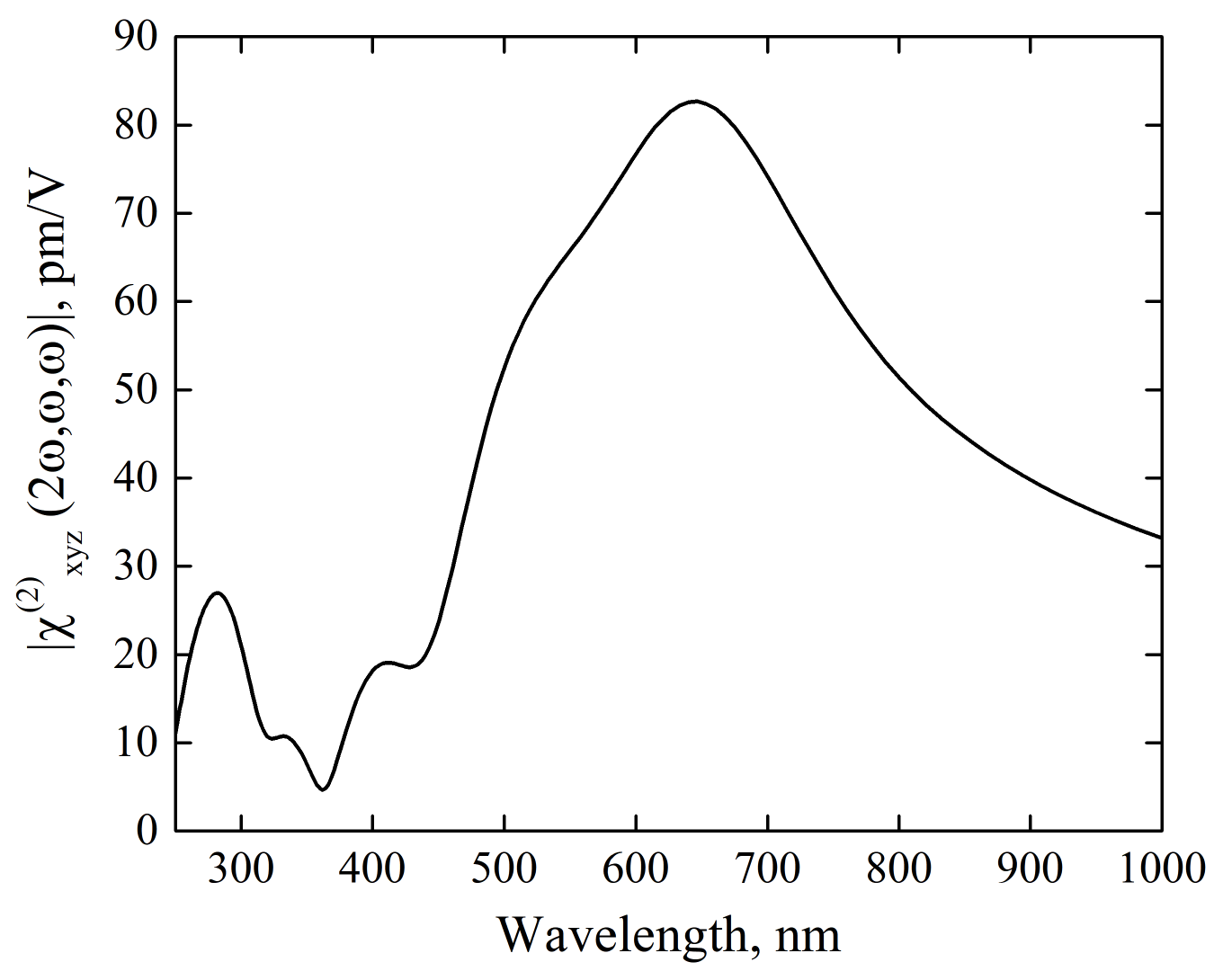}
\caption{\label{fig:nonlinear} Theoretical prediction of the SHG spectrum of
\ce{Y4H10O} in terms of the spectral behavior of the absolute value of 
the second-order susceptibility $\chi_{xyz}(2\omega,\omega,\omega)$. }
\end{figure}

Our next result follows from the analysis of whether the nonlinear optical effect would be strong enough in crystalline \ce{Y4H10O}. 
Because this material has no center of inversion, the following two features of the electron band picture determined its choice for study: 
the band gap lying in the visible light region, and 
higher electron polarizability compared to \ce{YHO}.
In the context of optical nonlinearity,
the focus was to clarify whether the electronic structure is capable of providing such an asymmetry of electronic transitions that could cause a high level of second-order polarization response to the electric field of the incident light wave.
Shown in Figure~\ref{fig:nonlinear} is the theoretical prediction for 
the independent component of the second harmonic generation (SHG)
susceptibility tensor $\chi_{xyz}(2\omega,\omega,\omega)$ calculated 
for the non-centrosymmetric cubic phase of \ce{Y4H10O}.
It can be seen that \ce{Y4H10O} exhibits a relatively large SHG coefficient for a relatively wide transparency range of $500$-$900$~nm.
A theoretical estimate for the light excitation energies corresponding to wavelengths of $281$ and $647$~nm yielded the values $\chi_{xyz}(2\omega)\,=\,27.0$ and $\chi_{xyz}(2\omega)\,=\,82.7$~pm/V, respectively.
These values are close to the nonlinear optical susceptibilities typical for perovskite ferroelectrics with strong optical nonlinearities~\cite{ATUCHIN2004411}.
Moreover, if an optical multilayer system uses a \ce{Y4H10O} layer, such as the one shown in Figure~\ref{fig:Fig_4a}, one can assume a possible enhancement of the nonlinear optical response in this system.
%
\section{\label{sec:level4}Conclusion}
Recent results from the research of oxyhydrides shed light on the fundamental issue of how the material properties of oxyhydride systems are related to their structural versatility.
The aim of the present work was to perform a computationally-oriented constructive analysis to understand how partial oxidation, stoichiometric effects, and differences in structure might allow systematic control of the optical properties of yttrium oxyhydrides.
The main idea was to obtain useful information that would allow us to describe the separation of different oxidation levels during synthesis and therefore would be helpful in creating a stoichiometric compound with a given amount of oxygen.
In this way, the functionality and efficiency of oxyhydride-based systems could be expanded for possible use in new electromechanical and optoelectronic devices.
We chose two stable modifications, \ce{Y4H10O} and \ce{YHO}, as effective models corresponding to structures with low and high amounts of chemically bound oxygen,
respectively.
To evaluate the important differences in the characteristics of the two models, we performed the simulations of structure-properties relationships by means of DFT calculations.
We found that the special role of oxygen is that its incorporation in stoichiometric amounts greatly varies electronic and optical properties.
Analysis of changes in the optical descriptors allowed us to propose an open configuration scheme for the design of a multilayer composite optical system, permitting wide variation of its spectral behavior through a matching of suitable structural elements.
Our theoretical estimates of 2nd order nonlinear susceptibility showed that noncentrosymmetric \ce{Y4H10O} may have a large optical nonlinearity potential.

\begin{figure*}[b]
\centering
     \includegraphics[width=1.0\textwidth]{Figure_01.pdf}
\caption{\label{fig:Comb_optics} Spectral behavior of optical characteristics  
of yttrium oxyhydrides, \ce{Y4H10O} and \ce{YHO}, simulated in terms of 
complex dielectric function $\epsilon$, refractive index $n$, 
reflectance $R$ and transmittance $T$ spectra.}
\end{figure*}

\section*{SUPPORTING INFORMATION}
Data details on material properties are presented in Supporting Information.

\section*{ACKNOWLEDGEMENTS}
E.S. acknowledges the support from the European Regional Development Fund.
A.P. was supported by the Estonian Research Council grant PRG 347.
S.Z.K. has received funding from the Research Council of Norway through 
project 309827. Computational work has been performed by using 
the Norwegian NOTUR supercomputing facilities through project nn4608k.

\section*{CONFLICT OF INTEREST}
The authors declare no conflict of interest.

\printendnotes
\bibliography{optics_manuscript.bib}


\end{document}